\documentclass{osa-article}

\journal{oe}

\articletype{Research Article}

\usepackage{lineno}
\nolinenumbers

\begin{document}

\title{Fabrication of low-loss lithium niobate on insulator waveguides on the wafer scale}

\author{Mohammadreza Younesi,\authormark{1,*} Thomas Käsebier,\authormark{2} Ilia Elmanov,\authormark{1} Yang-Teng Li,\authormark{1} Pawan Kumar,\authormark{1} Reinhard Geiss,\authormark{3} Thomas Siefke,\authormark{2,3} Falk Eilenberger,\authormark{2,3} Frank Setzpfandt,\authormark{1,3} Uwe Zeitner,\authormark{3,4} and Thomas Pertsch\authormark{1,3}}

\address{\authormark{1}Abbe Center of Photonics, Friedrich Schiller University, Albert-Einstein-Str. 6, 07745 Jena, Germany\\
\authormark{2}Institute of Applied Physics, Friedrich Schiller University, Albert-Einstein-Str. 15, 07745 Jena, Germany\\
\authormark{3}Fraunhofer Institute for Applied Optics and Precision Engineering IOF, Albert-Einstein-Str. 7, 07745 Jena, Germany\\
\authormark{4}Department of Applied Sciences and Mechatronics, Hochschule München, Lothstr. 34, 80335 Munich, Germany}

\email{\authormark{*}mohammadreza.younesi@uni-jena.de} 



\begin{abstract}
We report on the wafer scale fabrication of single mode low-loss lithium niobate on insulator waveguides utilizing a chemically amplified resist and an optimized dry etching method. The fabricated single mode waveguides are free of residuals and re-deposition, with measured losses for straight waveguides around 2~dB/m (0.02~dB/cm). We present on a method offering advantages for large-scale production due to its cost-effectiveness, faster writing time, and simplified processes. This work holds promise for advancing integrated photonics and optical communication technologies.
\end{abstract}

\section{Introduction}
Lithium niobate on insulator (LNOI) is a widely-used substrate for different applications in nonlinear optics, electro-optics, and acousto-optics due to its versatile properties and large refractive index contrast, which makes it a promising platform for photonic integrated circuits (PICs)\cite{Boes:18, Chen:22}. Such PICs are ideal candidates to enable the second quantum revolution by facilitating the control of quantum states on a large scale platform \cite{Lu:21}. LNOI-based PICs have applications beyond quantum technologies, such as coherent communication systems and ultrafast signal processing \cite{Xu:20,Wang:18,Feng:24}. For the continued development of LNOI PICs, a fabrication scheme is required to enable (i) low loss waveguides, ensuring their practicality and scalability in real-world applications, (ii) cost effectiveness, (iii) high flexibility in fabrication of different structures, and (iv) wafer scale operation. 

Among various fabrication techniques, electron beam lithography (EBL) is a prevalent method for realization of PICs, where an electron beam defines the pattern of waveguides and other structures in a sensitive electron beam (e-beam) resist, followed by different pattern transfer processes to etch the LN layer. Achieving flexibility is feasible through high throughput character projection EBL (CP-EBL) which is a compatible method with the processes from Semiconductor Equipment and Materials International (SEMI). In this method, pre-defined cell structures are projected onto a substrate using an electron beam, enabling precise patterning at the nanoscale \cite{Haedrich:22}. Furthermore, since writing speed is up to two orders of magnitude quicker than conventional Gaussian EBL, this provides a good balance between cost efficiency and flexibility. 

Waveguide losses stem from the intrinsic material absorption of LN and scattering from the waveguide. The scattering is increased by sidewall roughness of the waveguide. The sidewall roughness are the imperfection and irregularities in the shape of the waveguide, which leads to coupling between different guided modes of the waveguide and from the guided modes to radiation modes, increasing the propagation loss \cite{Poulton:06}. The sidewall roughness arises from the fabrication process, including imperfections in the shape of the mask used for the etching, anisotropy in the etching processes, and the material quality of LN. Controlling and reducing the sidewall roughness results in high quality waveguides and low propagation loss.

Many different methods were previously reported for realizing high quality and low loss waveguides in LNOI using EBL. The main methods with the lowest propagation losses which are widely used are chemo-mechanical polishing and dry etching. For the chemo-mechanical polishing technique, the desired structures are patterned as a chromium (Cr) mask on the surface of LN, and the pattern is transferred into LNOI using a wafer polishing machine. This method has been used for fabrication of low-loss structures with propagation loss of around 1~dB/m \cite{Li:23,Luo:21,Wu:18,Gao:23,Lin:22,Wang:19}. However this method is not compatible with standard fabrication processes and the fabricated waveguides have a large sidewall angle preventing high integration density.

Alternatively, nanostructuring of LN through utilization of a patterned mask on its surface, followed by transferring the pattern into LN using dry etching techniques shows promise for fabricating high-quality LNOI waveguides. Extensive research employing various etching methods and diverse mask materials has been conducted. Among the notable achievements, Zhang et al. demonstrated exceptional results by fabricating low-loss LNOI waveguides, achieving a propagation loss of 2.7 dB/m \cite{Zhang:17}. In this well-established fabrication method, the waveguides are fabricated using a thick hydrogen silsesquioxane (HSQ) resist layer, directly as the mask for etching into LN\cite{Luo:21,Ansari:22,Li:21,Luo:22}. This resist can provide high resolution and thick mask with a good selectivity and it can be used for directly etching into LN. Although HSQ offers a very high resolution, the required dose and thus writing time for pattering the resist is prohibitive for wafer scale applications. These properties make HSQ disadvantageous for volume production.

Compared to HSQ, chemically amplified resist (CAR) needs smaller dose and writing time, since it contains a photochemical acid generator (PAG), which acts as catalyst in the exposure process to increase the sensitivity of the resist\cite{Ito:2005}. The primary drawback of many CARs is their limited selectivity, necessitating the use of an intermediate hard mask. Previously many different research groups have published their results in realization of low-loss structures using different masks such as SiO\textsubscript{2} and Cr\cite{Krasnokutska:18,Luke:20,Bahadori:19,Li:18,Lin:21}. The reported waveguides have loss in the range of 20 dB/m. Even though the measured losses are comparable to the best losses reported, the process can be improved, further reducing the waveguide losses. Here we enhance this process, reducing the waveguide losses comparable to those fabricated using the CMP technique.

In this paper, we propose a fabrication method relying on a CAR with a minimized e-beam exposure dose, an intermediate hard mask and a refined dry etching processes for realization of low loss waveguides without using HSQ resist. First, the fabrication route is demonstrated in detail and afterwards loss measurement results from racetrack resonators are presented. In the final part, we calculate the propagation loss of the straight waveguide using numerical calculations.

\section{Fabrication}
Utilizing CARs and an intermediate hard mask is a technology more suited to high-volume production\cite{Ikeno:16} due to its lower exposure dose value and writing time. Among different CAR resists, we use FEP~171 with a dose value of 10~µC/cm\textsuperscript{2}, which is much smaller compared to the dose value needed for patterning HSQ resist ($\sim$5000~µC/cm\textsuperscript{2}). 

\begin{figure}[htbp]
\begin{center}
\includegraphics[width=7.6cm]{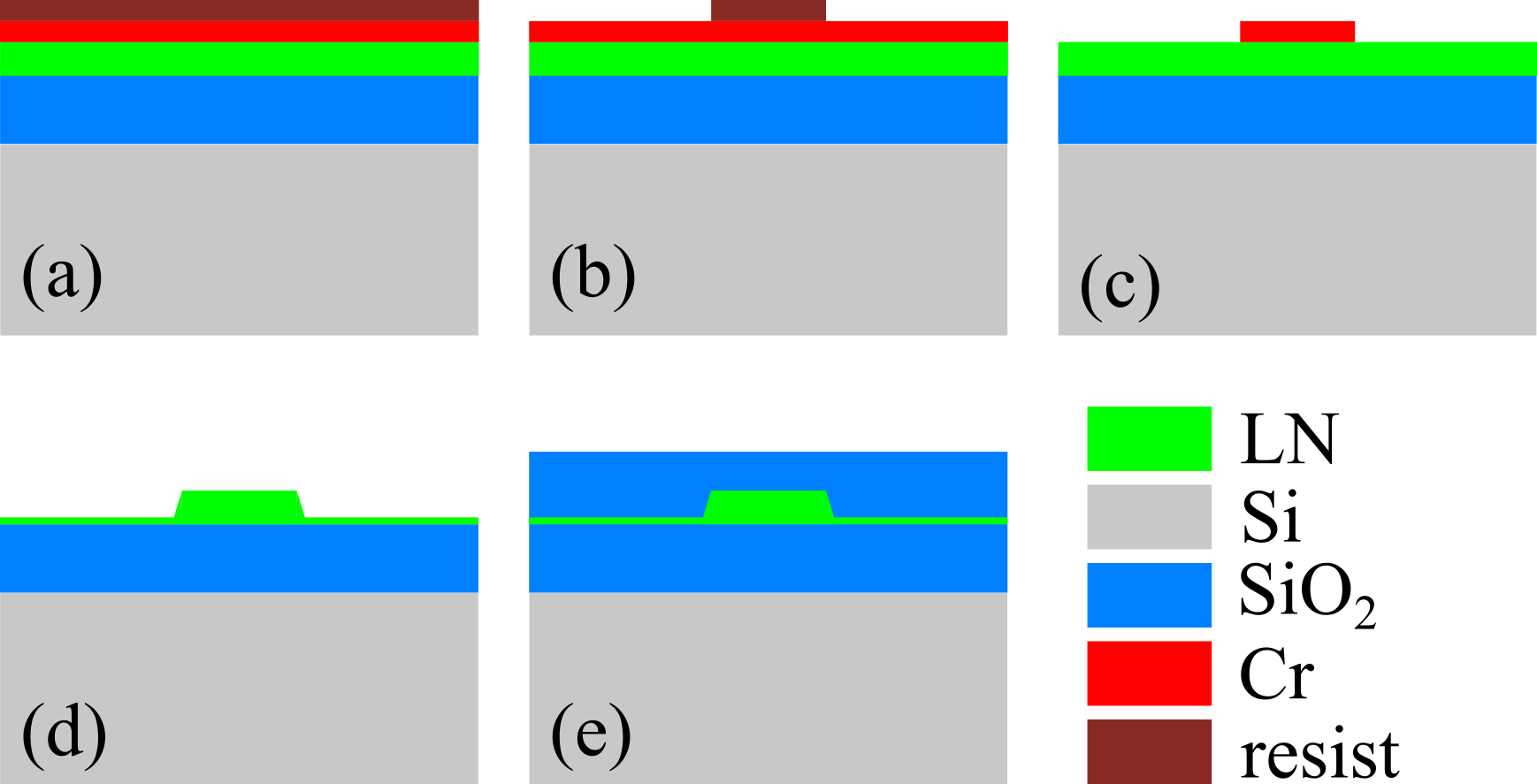}
\end{center}
\caption{Fabrication steps. (a)~deposition of Cr layer and resist (b)~patterning the resist (c)~RIE etching of Cr layer (d)~ICP-RIE etching of LN (e) annealing and cladding the structures with a layer of SiO\textsubscript{2}.}\label{fig:fabrication_steps}
\end{figure}

The fabrication method we follow for the realization of low-loss waveguides is to use an intermediate Cr hard mask. The complete process is shown in Fig. \ref{fig:fabrication_steps}.
The substrate used in this work consists of a 600-nm-thick LN on top of a layer of silicon dioxide ({SiO}\textsubscript{2}) with a thickness of 2~µm and a 525-µm-thick Si layer for handling (NANOLN, Co.). Our samples are in the form of 4-inch wafers. 
Firstly, the sample is covered with a 120-nm-thick Cr layer (ion beam deposition, Oxford Ionfab 300+ LC), followed by spin coating of FEP~171 resist layer with a thickness of 300~nm (Fig.~\ref{fig:fabrication_steps} (a)). This resist thickness is enough for etching 120~nm of Cr and transferring the pattern into the hard mask. In the next step (Fig.~\ref{fig:fabrication_steps}(b)), the resist is exposed by a variable shaped beam (VSB) electron beam writer (Vistec SB350 OS) with multi-pass writing mode \cite{Haedrich:22} and developed to define the structures.  Then, the pattern is transferred into the hard mask (Fig.~\ref{fig:fabrication_steps} (c)) using the reactive ion etching method (Sentech SI-591), followed by etching into LN layer (Fig.~\ref{fig:fabrication_steps} (d)) by inductively-coupled plasma reactive ion etching (ICP-RIE) with CHF\textsubscript{3} gas (Sentech SI-500 C). 

One of the main challenges we faced for nanostructuring of LN is the presence of an amorphous layer by re-deposition of the material removed from the LN substrate during the etching process which sticks to the fabricated nanostructures\cite{Kaufmann:23}. Several methods such as chemical cleaning or ion milling\cite{Ulliac:16, Escale_thesis} for removing the re-deposition layers have been introduced, however,  removing this layer could cause additional sidewall roughness on the waveguides. The best approach to address this problem is to optimize the fabrication process itself and avoid the deposition of the amorphous layer in the first place. Previously it has been shown that for etching with ICP-RIE, by optimizing the etching parameters such as chamber pressure and DC bias, the re-deposition layer can be controlled\cite{Kaufmann:23}. In our fabrication development, we optimized the etching conditions for patterning LN to ensure the fabrication of high quality structures without any re-deposition and residuals.

Afterwards, the sample is cleaned and prepared for measurement. We annealed the sample in an oxygen (O\textsubscript{2}) atmosphere at room pressure, maintaining a temperature of 520°C for 2 hours. This procedure results in a reduction of the loss of waveguides by improving the crystallinity of the LN \cite{Ansari:22,Luo:21}. Finally, we apply an SiO\textsubscript{2} cladding layer (Plasma-enhanced chemical vapor deposition) to the sample to enhance mode field distribution and provide additional protection (Fig.~\ref{fig:fabrication_steps} (e)).

\section{Discussion and Results}

\begin{figure}[ht!]
\centering\includegraphics[width=6cm]{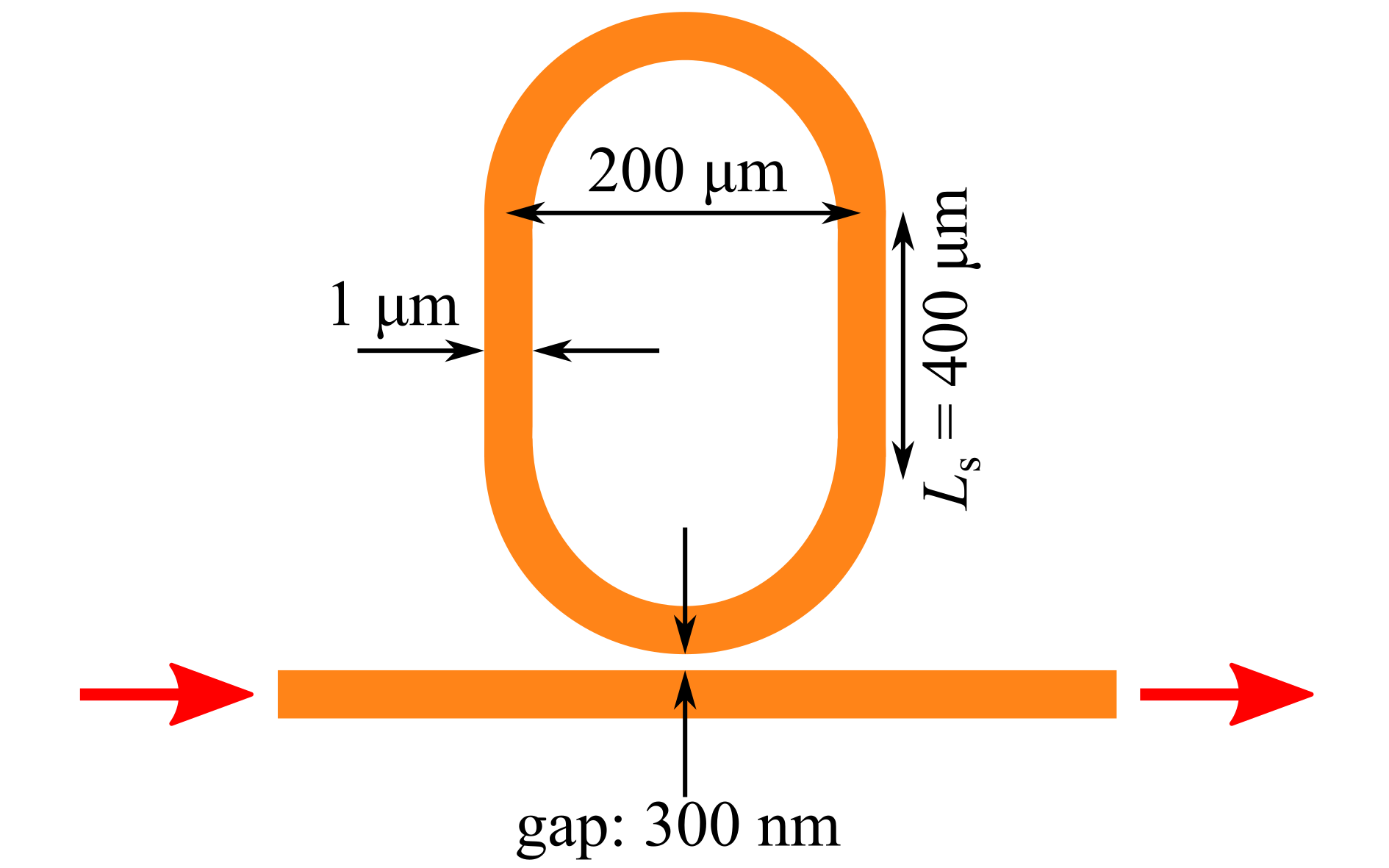}
\caption{Schematic of the racetrack resonator and waveguide.}
\label{fig:racetrack}
\end{figure}

To measure the propagation losses of the fabricated structures, we fabricated racetrack resonators. The schematic of our structure is illustrated in Fig.~\ref{fig:racetrack}. Light is coupled into the waveguide using the edge coupling method, and this light is then evanescently coupled into the racetrack resonator within the gap region. The transmitted light is collected with a lensed fiber, detected with a photodetector, and used for calculation of the waveguide loss in the resonator. In our structures, the LN layer is etched for 500~nm, the top width of the structures for the waveguide and racetrack resonators is 1~µm. The length of the straight waveguide part of the resonator is 400~µm, all bend radii are 100~µm, and the gap between the waveguide and resonator at the coupling region is 300~nm.

\begin{figure}[ht!]
\centering\includegraphics[width=13.5cm]{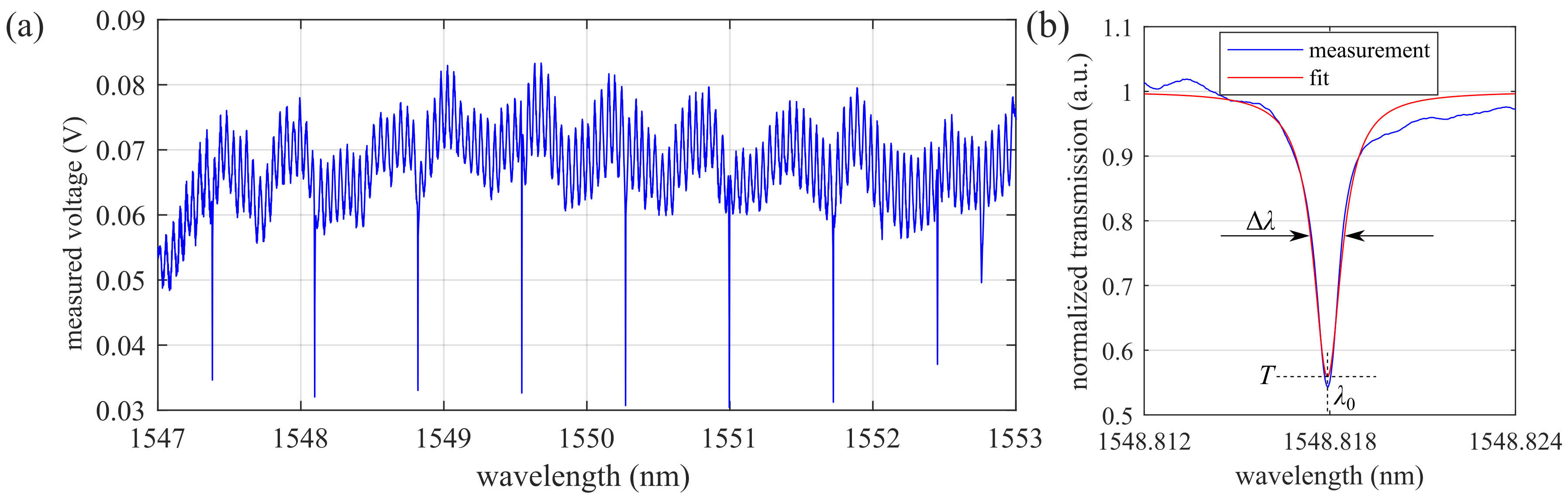}
\caption{(a)~Measured voltage corresponding to the transmitted signal of the resonator, in the range of 1547 to 1553~nm. (b)~The resonance at 1548.818~nm and the Lorentzian fit function. The plots are normalized to 1. (c)~Variation of two loss values ${l}\textsubscript{1,2}$ in the wavelength range of 1440 to 1640 nm.}
\label{fig:loss2}
\end{figure}

The total loss of the resonator stems from two effects: the scattering and absorption loss in the ring resonator (propagation loss) and the loss due to the coupling to the waveguide (coupling loss). To estimate the loss values, we performed transmission measurements of the racetrack resonator. The detected signal for the TE mode in the wavelength range of 1547 to 1553~nm is shown in Fig.~\ref{fig:loss2}(a). In the next step, the resonances in the spectrum are fitted using a Lorentzian function and normalized to the transmitted power outside of resonances. One of the resonances at the wavelength of 1548.818~nm and its fit is shown in Fig.~\ref{fig:loss2}(b). From the fitted data for the resonances, we calculated the value of the loaded quality factor ($Q$) and the group index ($n\textsubscript{g}$) of each resonance as
\begin{equation}
\label{eq:q}
Q =\frac{\lambda\textsubscript{0}}{\Delta\lambda}
\end{equation}
and
\begin{equation}
\label{eq:ng}
n\textsubscript{g} =\frac{\lambda\textsubscript{0}\textsuperscript{2}}{F\times{L\textsubscript{tot}}},
\end{equation}
where $\lambda\textsubscript{0}$ and $\Delta\lambda$ are the central wavelength and full width at half maximum of each resonance  (Fig.~\ref{fig:loss2}(b)), and  $F$ and $L\textsubscript{tot}$ are free spectral range of the resonances and total length of the resonator respectively. These values allow us to calculate two loss coefficients $\alpha\textsubscript{1,2}$ and the corresponding loss values $l\textsubscript{1,2}$ as 
\begin{equation}
\label{eq:alpha}
{\alpha}\textsubscript{1,2} =\frac{\pi n\textsubscript{g}}{Q\lambda}(1\pm\sqrt{T})
\end{equation}
and
\begin{equation}
\label{eq:loss}
{l}\textsubscript{1,2}=-10\log{(e^{-{\alpha}\textsubscript{1,2}\times 1[m]})},
\end{equation}
where $T$ is the minimal transmission of the normalized fit (Fig.~\ref{fig:loss2}(b)) \cite{Adar:91,An:18}. One of the calculated loss values relates to the propagation loss and the other one to the coupling loss of the racetrack resonator . If these two values are equal, the propagation and coupling loss values are the same and the resonator works in the critical coupling regime, which is not always the case.  For the measured wavelength range shown in Fig.~\ref{fig:loss2} (a), the averaged Q is $(1.34 \pm 0.24) \times 10^6$, $n\textsubscript{g}$ is $2.317\pm0.005$, and the corresponding loss values are ${l}\textsubscript{1}=4.76\pm0.34$~dB/m and ${l}\textsubscript{2}=26.7\pm 5.2$~dB/m. This means our resonator is not working in the critical coupling regime.

In order to determine which one of the two measured losses represents the actual propagation loss of the waveguide and which one the coupling loss, we carried out a simulation using the dimensions of the measured racetrack resonator. The coupling loss determined from the simulation is 28.97~dB/m which is close to the measured loss ${l}\textsubscript{2}$.
It is safe to assume that ${l}\textsubscript{1}$ is the propagation loss of the waveguide, while ${l}\textsubscript{2}$ represents the coupling loss. In other words, our resonator is working in the strong coupling regime and the propagation loss ($l\textsubscript{p}$ ) and coupling loss ($l\textsubscript{c}$) values in the wavelength range of 1447 to 1553~nm for TE mode are around 5~dB/m and 27~dB/m respectively. We also performed similar measurements and analysis for the TM mode of the resonator and the measured propagation loss for the same wavelength range for TM mode is around 8~dB/m. The propagation loss has two different contributions: straight waveguide loss and bend loss. We expect the propagation loss for the straight waveguide to be smaller than the measured value for the entire racetrack resonator, hence we carried out the following calculations to find the actual loss of the straight waveguide.

The calculated propagation loss values explained previously are under the assumption of having identical propagation loss for straight and curved parts of the racetrack resonator. This assumption is generally not true because the losses induced by waveguide bending are not considered. Here, we separate these two losses and calculate the propagation loss of the straight waveguide.

Since the measured resonator is working in the strong coupling regime, the propagation ($\alpha\textsubscript{p}$) and coupling ($\alpha\textsubscript{c}$) loss coefficients can be calculated from Eq. (\ref{eq:alpha}) as
\begin{equation}
\label{eq:alphap}
\alpha\textsubscript{p} =\frac{\pi n\textsubscript{g}}{Q\lambda}(1-\sqrt{T})
\end{equation}
and
\begin{equation}
\label{eq:alphac}
{\alpha}\textsubscript{c} =\frac{\pi n\textsubscript{g}}{Q\lambda}(1+\sqrt{T}).
\end{equation}
Under the assumption of small losses and small coupling one can get equations to calculate coupling and propagation losses, constructing a system of two equations \cite{Adar:91}:
\begin{equation}
\label{eq:q1}
\frac{1}{Q} = \frac{\alpha\textsubscript{p} \lambda}{2\pi{n}\textsubscript{g}} + \frac{\alpha\textsubscript{c} \lambda}{2\pi{n}\textsubscript{g}}
\end{equation}
\begin{equation}
\label{eq:ratio}
\frac{\alpha\textsubscript{c}}{\alpha\textsubscript{p}} = \frac{1+ \sqrt{T}}{1-\sqrt{T}},
\end{equation}
where $\lambda$ is the central wavelength of the measured bandwidth (here it is 1550 nm). Separating the loss coefficients for the straight ($\alpha\textsubscript{s}$) and curved ($\alpha\textsubscript{r}$) parts of the resonator with $\alpha\textsubscript{p}=\alpha\textsubscript{s}+\alpha\textsubscript{r}$, the equations change as 
\begin{equation}
\label{eq:q2}
\frac{1}{Q} = (\alpha\textsubscript{r}+\alpha\textsubscript{s}+\alpha\textsubscript{c})\frac{\lambda}{2\pi{n}\textsubscript{g}}
\end{equation}
and
\begin{equation}
\label{eq:ratio2}
\frac{\alpha\textsubscript{c}L\textsubscript{tot}}{\alpha\textsubscript{r}2\pi R + \alpha\textsubscript{s}2L\textsubscript{s}} = \frac{1+\sqrt{T}}{1-\sqrt{T}},
\end{equation}
where $R$ is the bending radius and $L\textsubscript{s}$ is the length of straight parts of the resonator 
(as it is shown in Fig.~\ref{fig:racetrack}).
These two equations contain three unknown values and solving it requires an additional equation. To solve this problem, we compare the measurement results for two similar racetrack resonators which are only different in the length of straight waveguide $L\textsubscript{s}$. We carried out measurements for two resonators with $L\textsubscript{s}$ of 100 and 500~µm in the range of 1447 to 1553~nm. Since these two resonators have only different $L\textsubscript{s}$ compared to previously measured resonator, we assume that both resonators are working in the strong coupling regime and they have equal $\alpha\textsubscript{r}$ and $\alpha\textsubscript{s}$ and different $\alpha\textsubscript{c}$. This adds two more equations and one more unknown variable.
Labeling the corresponding parameters for resonators with $L\textsubscript{s}$ of 100 and 500~µm respectively by I and II, the system of equations is now
\begin{equation}
    \label{eq:ringcoup}
    \scalebox{1}{
    $\begin{pmatrix}
        (1- \sqrt{T\textsuperscript{I}})L\rlap{\textsuperscript{I\textsuperscript{I}}}\textsubscript{tot} & -(1+\sqrt{T\textsuperscript{I}})2\pi R & -(1+ \sqrt{T\textsuperscript{I}})2L\rlap{\textsuperscript{I}}\textsubscript{s} & 0 \\
        1 & 1 & 1 & 0 \\
        0 & -(1+\sqrt{T\textsuperscript{II}})2\pi R & -(1+ \sqrt{T\textsuperscript{II}})2L\rlap{\textsuperscript{II}}\textsubscript{s} & (1- \sqrt{T\textsuperscript{II}})L\rlap{\textsuperscript{II}}\textsubscript{tot} \\
        0 & 1 & 1 & 1
        \end{pmatrix}$
    }
    \scalebox{1}{
        $\begin{pmatrix}
        \alpha\rlap{\textsuperscript{I}}\textsubscript{c} \\
\alpha\textsubscript{r} \\
\alpha\textsubscript{s} \\ \alpha\rlap{\textsuperscript{II}}\textsubscript{c}
        \end{pmatrix}$
    }    
    =
    \scalebox{1}{
        $\begin{pmatrix}
        0 \\
        \frac{2\pi{n}\textsubscript{g}}{Q\textsuperscript{I}\lambda} \\
        0 \\
        \frac{2\pi{n}\textsubscript{g}}{Q\textsuperscript{II}\lambda}
        \end{pmatrix}$
    }.
\end{equation}

Solving the equations, we calculate interval values for $\alpha\rlap{\textsuperscript{I}}\textsubscript{c}$, $\alpha\rlap{\textsuperscript{II}}\textsubscript{c}$, $\alpha\textsubscript{r}$, and
$\alpha\textsubscript{s}$ with 90\% confidence intervals and then using an exponential equation similar to Eq. \ref{eq:loss} we calculate the corresponding loss value
$l\rlap{\textsuperscript{I}}\textsubscript{c}=24.23\pm7.70$~dB/m, 
$l\rlap{\textsuperscript{II}}\textsubscript{c}=22.02\pm4.83$~dB/m, 
$l\textsubscript{r}=11.37\pm3.37$~dB/m, and
$l\textsubscript{s}=0.37\pm1.74$~dB/m.
This shows that the propagation loss for TE mode in the straight waveguide is highly likely less than 2~dB/m, with good potential to be well below 1~dB/m.
\section{Conclusion and outlook}
In this work, the fabrication of LNOI waveguides using dry etching is investigated. With the focus on smooth sidewalls for low loss, a suitable process based on the ICP-RIE dry etching method is found. We used FEP~171 which has a small dose value compared to other resists and is a good candidate for volume production. We optimized the fabrication process to avoid the residuals and re-deposition layer and achieved high-quality structures. Our measured overall propagation loss for the fabricated racetrack resonators for the TE mode is around 5~dB/m and for the TM mode is around 8~dB/m, which is in the range of the best loss reported in the literature. Further analysis shows that the propagation loss of straight waveguide section of the resonator is smaller and in the range of 2~dB/m.

\section{Acknowledgement}
The sample fabrication within this work was carried out by the microstructure technology team at IAP Jena.
The authors would like to thank them for providing the fabrication facilities, carrying out processes and providing support.

\section{Funding}
German Federal Ministry of Education and Research (BMBF) under the project identiﬁers 13N14877 (QuantIm4Life), 13N16108 (PhoQuant); German Research Foundation (DFG) under the project identiﬁers PE 1524/13-1 (NanoPair), 398816777-SFB 1,375 (NOA); Thuringian Ministry for Economic Affairs, Science and Digital Society under the project identiﬁer 2021 FGI 0043 (Quantum Hub Thuringia); Carl-Zeiss-Foundation (CZS Center QPhoton).

\section{Disclosures}
The authors declare no conflicts of interest. 

\section{Data availability}
Data underlying the results presented in this paper are not publicly available at this time but may be obtained from the authors upon reasonable request.

\bibliography{references}
\end{document}